# Violation of the scaling relation and non-Markovian nature of earthquake aftershocks*

Sumiyoshi Abe[1,2] and Norikazu Suzuki[3]

[1] *Department of Physical Engineering, Mie University, Mie 514-8507, Japan*

[2] *Institut Supérieur des Matériaux et Mécaniques Avancés, 44 F. A. Bartholdi, 72000 Le Mans, France*

[3] *College of Science and Technology, Nihon University, Chiba 274-8501, Japan*

**Abstract**   The statistical properties of earthquake aftershocks are studied. The scaling relation for the exponents of the Omori law and the power-law calm time distribution (i.e., the interoccurrence time distribution), which is valid if a sequence of aftershocks is a singular Markovian process, is carefully examined. Data analysis shows significant violation of the scaling relation, implying the non-Markovian nature of aftershocks.

PACS numbers:    91.30.Dk, 02.50.Ey, 05.45.Tp, 05.40.-a

───────────────────────────────────────────────

*) Dedicated to François Bardou (1968-2006).

Markovian stochastic process is ubiquitous in nature. A celebrated example is the Brownian motion and its associated normal diffusion phenomenon. An associated system has short-term memory. More precisely, in a Markovian process, transition between states is essentially due to each elementary fluctuation that can be treated locally. On the other hand, the non-Markovian nature signals a certain level of complexity of a system, which may exhibit correlated anomalous diffusion (see Refs. [1,2], for example). There, memory is long-term and accordingly local treatment is not possible. It is also of interest to see these points in analogy with nonlinear dynamics. For example, the logistic map in its chaotic regime (where the Lyapunov exponent is positive) quickly forgets its initial condition in the course of time evolution, whereas at the edge of chaos (where the Lyapunov exponent vanishes) the system remembers its initial condition even after long duration of time [3].

In physics of earthquakes, a number of efforts have been devoted to modeling seismicity based on the stochastic approach [4]. An important question there is if seismicity is non-Markovian, and this is generically a nontrivial issue.

A shallow strong earthquake tends to reorganize the stress distribution at faults. Accordingly, such an earthquake is usually followed by a number of aftershocks. A part of the seismic time series, in which the Omori law for temporal pattern of aftershocks holds, is refereed to here as *Omori regime*. A process in an Omori regime is highly nonstationary and event-event correlation decays *slowly*, exhibiting the aging phenomenon and obeying the scaling law [5]. These features are also observed in numerical analysis of the coherent noise model [6]. The results indicate that the mechanism governing aftershocks may be a type of glassy dynamics.

In this paper, we would like to study the non-Markovian nature of earthquake aftershocks by examining a scaling relation which has to be satisfied by a class of singular Markovian jump processes. We find that the scaling relation is in fact significantly violated by real aftershocks. This finding has obvious importance as a quantitative characterization of the non-Markov nature of seismicity.

First, let us recapitulate the scaling relation in a singular Markov process. This relation was originally discussed in the context of Lévy statistics in laser cooling of atoms [7] and was later proved mathematically [8]. It is concerned with two quantities in point processes. One is the calm time distribution (i.e., the interoccurrence time distribution), $P$, which is the distribution of time interval between two successive events, and the other is the number of events found in the time interval $[0, t]$, $N(t)$. If the process is Markovian, then holds the following equation [7,8]:

$$S(t) = P(t) + \int_0^t dt'\ P(t-t')\ S(t'), \tag{1}$$

which is derived from the Kolmogorov forward equation. Here, $S(t) \equiv [N(t+\Delta t) - N(t)]/\Delta t$, which is the mean density of events at time $t$. The Laplace transformation of Eq. (1) yields

$$£[S](s) = \frac{£[P](s)}{1 - £[P](s)} \tag{2}$$

with the notation $£[f](s) = \int_0^\infty dt\ e^{-st}\ f(t)$.

Consider the case when both $P$ and $S$ are of the power-law type:

$$P(\tau) \sim \frac{1}{\tau^{1+\mu}} \quad (\tau \to \infty), \tag{3}$$

$$S(t) \sim \frac{1}{t^p} \quad (t \to \infty). \tag{4}$$

If the exponents, $\mu$ and $p$, are in the ranges

$$0 < \mu < 1, \qquad 0 < p < 1, \tag{5}$$

then the Laplace transformations of $P$ and $S$ behave as

$$\pounds[P](s) \sim 1 - \alpha s^\mu \quad (s \to 0), \tag{6}$$

$$\pounds[S](s) \sim \frac{1}{s^{1-p}} \quad (s \to 0), \tag{7}$$

respectively, where $\alpha$ is a positive constant. The range of $p$ in Eq. (5) means that the total number of events, $N(\infty)\,(=\pounds[S](0))$, is divergent in an idealized situation. From Eq. (2), it follows that

$$p + \mu = 1, \tag{8}$$

which is a scaling relation to be fulfilled by Markovian processes with $P$ and $S$ of the forms in Eqs. (3) and (4).

In seismology, $S$ is directly related to the Omori law [9], which states that frequency of aftershocks obeys

$$S(t) \equiv \frac{d N(t)}{d t} = \frac{A}{(1 + t / t_0)^p}, \tag{9}$$

where $A$ and $t_0$ are positive constant. It is an empirical fact that the exponent, $p$, actually ranges between 0.5 and 1.5. Therefore, to examine the scaling relation in Eq. (8) it is necessary to choose Omori regimes with exponent $p$ in the range in Eq. (5).

On the other hand, $P$ describes the calm time distribution. In Ref. [10], it has been reported that it decays as a power law (see also Ref. [11]). However, the analysis performed there was not restricted to a time interval of aftershocks alone but a long interval of the seismic time series. Therefore, it is also necessary to reanalyze the calm time distribution by limiting ourselves to the Omori regimes.

We have performed these analyses by employing the seismic data taken from California, which is currently available at http://www.data.scec.org/. In particular, we have focused our attention to the following two main shocks. (a) The Landers Earthquake with M7.3 occurred at 11:57:34.13 on June 28, 1992 ($34°12.00'$ N latitude, $116°26.22'$ W longitude, 0.97 km in depth). (b) The Hector Mine Earthquake with M7.1 occurred at 09:46:44.13 on October 16, 1999 ($34°35.64'$ N latitude, $116°16.26'$ W longitude, 0.02 km in depth). These main shocks are very shallow and thus were followed by swarms of aftershocks. As the Omori regimes, (a) 600 days are taken and the number of aftershocks contained is 34783, and (b) 370 days are taken and accordingly the number of aftershocks is 17368, respectively.

In Fig. 1, we present the plots of the mean density of events, $S$. There, we see that the Omori law in Eq. (4) holds well. [Rapid dropping due to smallness of the size of the Omori regime is observed in Fig. 1 (b). In fact, the Omori regime of (b) is much shorter than that of (a).] On the other hand, the plots of the calm time distribution, $P$, are given in Fig. 2. It is seen that the distribution decays as a power law, obeying Eq. (3).

However, a careful analysis is needed for determining the values of the exponent, $\mu$, since the scaling regions are not so wide. To evaluate the values of $p$ and $\mu$, we have used the method of maximum likelihood estimation [12].

The result is summarized in Table I. As can be seen, in both cases of the Landers Earthquake and the Hector Mine Earthquake, $p$ and $\mu$ satisfy the conditions in Eq. (5). As one appreciates, the scaling relation in Eq. (8) is significantly violated, showing how the Markovian nature of aftershocks is violated.

In conclusion, we have analyzed the statistical properties of the Omori regimes of the Landers Earthquake and the Hector Mine Earthquake and have shown that their calm time distributions decay as a power law. Then, we have examined the scaling relation, which is valid for a singular Markovian process, and have found that it is significantly violated. In this way, we have presented an evidence that aftershock sequences are non-Markovian.

**Acknowledgments**

One of the authors (S. A.) would like to thank the late François Bardou for many valuable discussions developed during 2003 and 2004, which have initiated the present study. The work of S. A. was supported in part by a Grant-in-Aid for Scientific Research from the Japan Society for the Promotion of Science.

———————————————

[1] R. Morgado, F. A. Oliveira, G. G. Batrouni, and A. Hansen,

Phys. Rev. Lett. **89**, 100601 (2002).

[2] J. D. Bao and Y. Z. Zhuo, Phys. Rev. Lett. **91**, 138104 (2003).

[3] A. Robledo, Physica A **370**, 449 (2006).

[4] S. Shlien and M. N. Toksöz, Geophysical Journal International **42**, 49 (1975);

   D. B. Rosenblatt and L. Knopoff, J. Nonlinear Sci. **1**, 279 (1991);

   R. Yulmetyev, F. Gafarov, P. Hänggi, R. Nigmatullin, and S. Kayumov,

   Phys. Rev. E **64**, 066132 (2001);

   E. E. Alvarez, Methodology and Computing in Applied Probability **7**, 119 (2005);

   J. E. Ebel, D. W. Chambers, A. L. Kafka, and J. A. Baglivo,

   Seism. Res. Lett. **78**, 57 (2007).

[5] S. Abe and N. Suzuki, Physica A **332**, 533 (2004).

[6] U. Tirnakli and S. Abe, Phys. Rev. E **70**, 056120 (2004).

[7] F. Bardou, J. P. Bouchaud, A. Aspect, and C. Cohen-Tannoudji,

   *Lévy Statistics and Laser Cooling* (Cambridge University Press, Cambridge,

   2002).

[8] O. E. Barndorff-Nielsen, F. E. Benth, and J. L. Jensen,

   Adv. Appl. Prob. **32**, 779 (2000).

[9] F. Omori, J. Coll. Sci. Imper. Univ. Tokyo **7**, 111 (1894);

   T. Utsu, Geophys. Mag. **30**, 521 (1961).

[10] S. Abe and N. Suzuki, Physica A **350**, 588 (2005).

[11] A. Corral, Phys. Rev. Lett. **92**, 108501 (2004).

[12] M. L. Goldstein, S. A. Morris, and G. G. Yen, Eur. Phys. J. B **41**, 255 (2004);

   M. E. J. Newman, Contemporary Physics **46**, 323 (2005).

# Figure and Table Captions

FIG. 1  The log-log plots of the mean density of events, *S*, in the Omori regimes of (a) the Landers Earthquake and (b) the Hector Mine earthquake. In both cases, the bin size for constructing the histograms is taken to be 1 day, here.

FIG. 2  The log-log plots of the calm time distributions, *P*, in the Omori regimes of (a) the Landers Earthquake and (b) the Hector Mine earthquake, corresponding those in Fig. 1. In both cases, the bin size for constructing the histograms is taken to be 20 [s], here.

Table I  The values of the exponents *p* and $\mu$ for some values of the bin size for constructing the histograms.

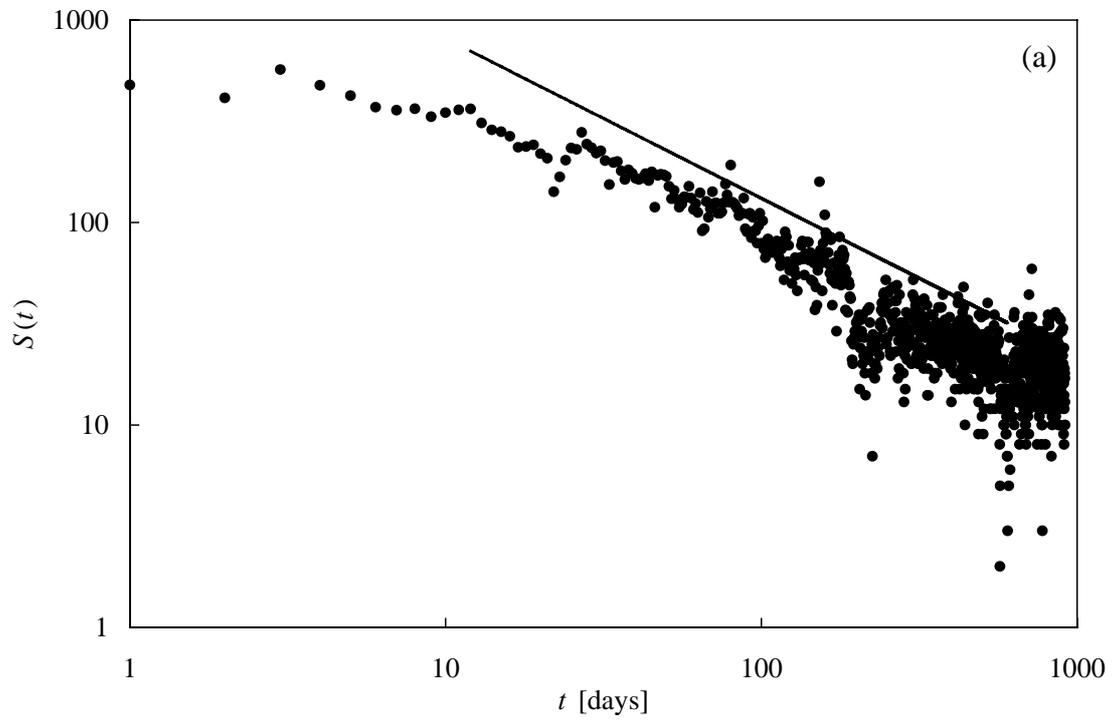

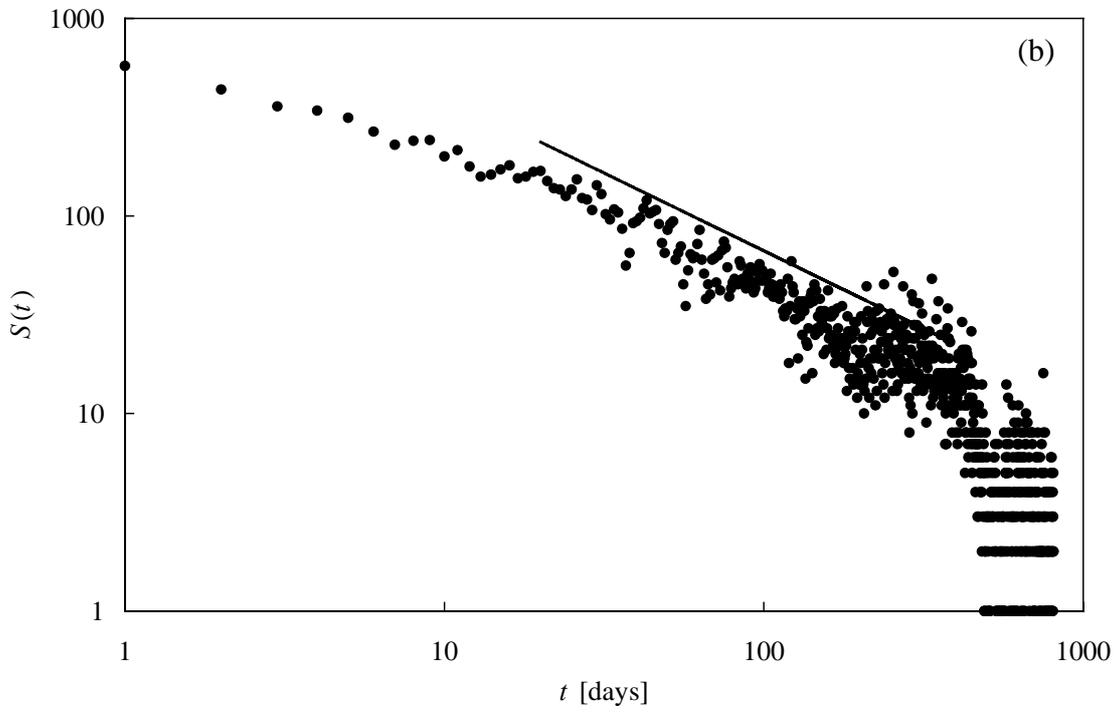

FIG. 1

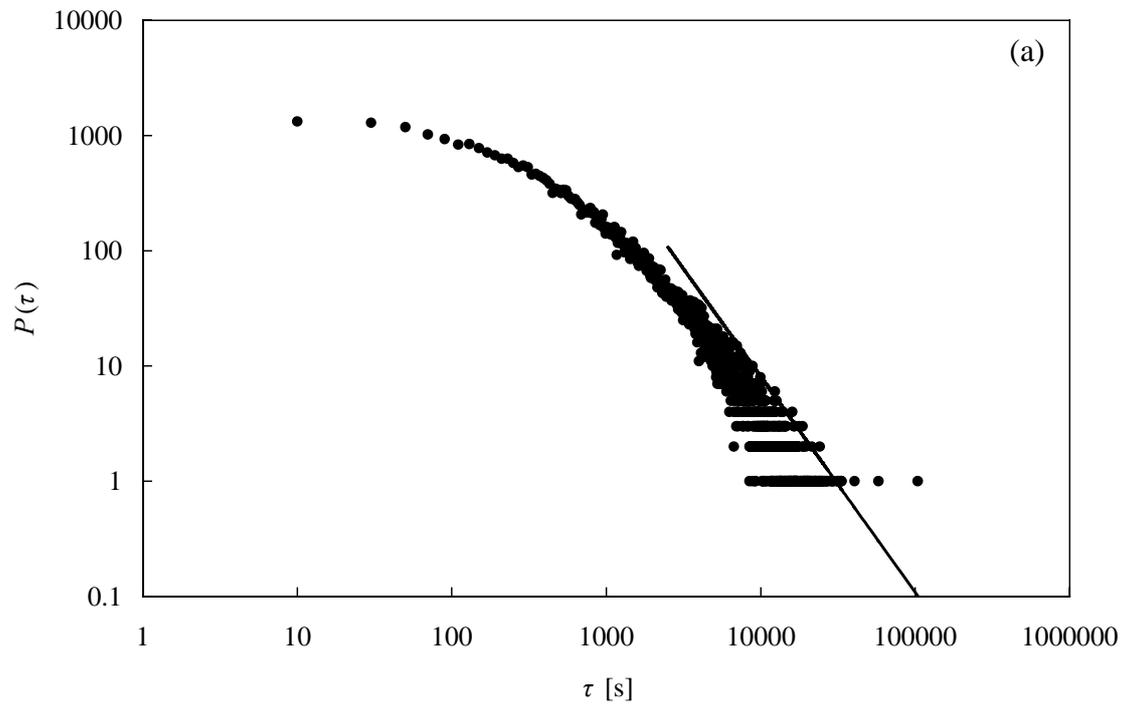

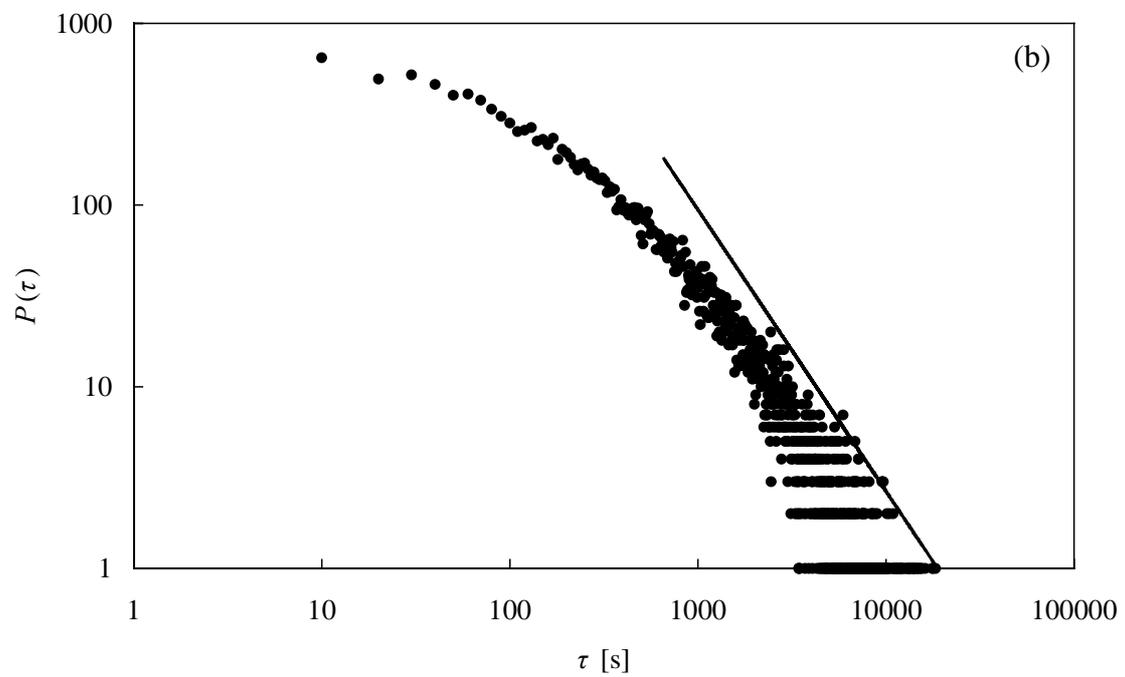

FIG. 2

|  | bin size [days] | $p$ | bin size [s] | $\mu$ | $p + \mu$ |
|---|---|---|---|---|---|
| Landers Earthquake | 0.5 | $0.788 \pm 0.004$ | 10 | $0.629 \pm 0.011$ | $1.417 \pm 0.015$ |
|  |  |  | 20 | $0.873 \pm 0.018$ | $1.661 \pm 0.021$ |
|  |  |  | 30 | $0.950 \pm 0.017$ | $1.738 \pm 0.021$ |
|  | 1 | $0.790 \pm 0.004$ | 10 | $0.629 \pm 0.011$ | $1.419 \pm 0.015$ |
|  |  |  | 20 | $0.873 \pm 0.018$ | $1.662 \pm 0.021$ |
|  |  |  | 30 | $0.950 \pm 0.017$ | $1.740 \pm 0.021$ |
|  | 1.5 | $0.793 \pm 0.004$ | 10 | $0.629 \pm 0.011$ | $1.422 \pm 0.015$ |
|  |  |  | 20 | $0.873 \pm 0.018$ | $1.665 \pm 0.021$ |
|  |  |  | 30 | $0.950 \pm 0.017$ | $1.743 \pm 0.021$ |
| Hector Mine Earthquake | 0.5 | $0.785 \pm 0.008$ | 10 | $0.408 \pm 0.014$ | $1.193 \pm 0.022$ |
|  |  |  | 20 | $0.550 \pm 0.013$ | $1.335 \pm 0.020$ |
|  |  |  | 30 | $0.618 \pm 0.013$ | $1.403 \pm 0.020$ |
|  | 1 | $0.787 \pm 0.007$ | 10 | $0.408 \pm 0.014$ | $1.195 \pm 0.022$ |
|  |  |  | 20 | $0.550 \pm 0.013$ | $1.337 \pm 0.020$ |
|  |  |  | 30 | $0.618 \pm 0.013$ | $1.405 \pm 0.020$ |
|  | 1.5 | $0.789 \pm 0.008$ | 10 | $0.408 \pm 0.014$ | $1.195 \pm 0.022$ |
|  |  |  | 20 | $0.550 \pm 0.013$ | $1.335 \pm 0.020$ |
|  |  |  | 30 | $0.618 \pm 0.013$ | $1.403 \pm 0.020$ |

Table I